\documentclass[12pt]{article}
\usepackage{amssymb,amsmath}
\usepackage{epsfig}
\newcommand{\sect}[1]{\setcounter{equation}{0}\section{#1}}

\def\N{{\mathcal N}}
\def\L{{\mathcal L}}

\def\O{{\mathcal O}}
\def\ds{\displaystyle}

\def\a{\alpha}
\def\b{\beta}

\def\g{\gamma}
\def\G{\Gamma}
\def\s{\sigma}
\def\t{\tilde}

\def\k{\kappa}
\def\f{\phi}
\def\vf{\varphi}

\def\th{\theta}

\def\o{\omega}
\def\p{\partial}

\def\axs{AdS_5\times S^5}
\newcommand{\eq}[1]{\begin{equation} #1 \end{equation}}
\newcommand{\al}[1]{\begin{align} #1 \end{align}}
\newcommand{\ml}[1]{\begin{multline} #1 \end{multline}}

\begin{document}

\begin{center}
{\bf{\Large Semiclassical Strings, Dipole Deformations \\
\ \\ of $\N=1$ SYM  and Decoupling of KK Modes}  \\
\vspace*{.35cm} }

\vspace*{1cm} N.P. Bobev${}^{\star\dagger}$, H. Dimov${}^{\ddag}$
and R.C. Rashkov${}^{\dagger}$\footnote{e-mail:
rash@phys.uni-sofia.bg; bobev@usc.edu}

\ \\
${}^{\star}$ \textit{Department of Physics and Astronomy ,
University of Southern California, Los Angeles, CA 90089-0484,
USA}

\ \\

${}^{\ddag}$ \textit{Department of Mathematics, University of
Chemical Technology and Metallurgy, 1756 Sofia, Bulgaria}

\ \\

${}^{\dagger}$ \textit{Department of Physics, Sofia University,
1164 Sofia, Bulgaria}

\end{center}

\vspace*{.8cm}

\begin{abstract}
In this paper we investigate the recently found $\g$-deformed
Maldacena-Nunez background by studying the behavior of different
semiclassical string configurations. This background is
conjectured to be dual to dipole deformations of $\N=1$ SYM. We
compare our results to those in the pure Maldacena-Nunez
background and show that the energies of our string configurations
are higher than in the undeformed background. Thinking in the
lines of (hep-th/0505100) we argue that this is an evidence for
better decoupling of the Kaluza-Klein modes from the pure SYM
theory excitations. Moreover we are able to find a limit of the
background in which the string energy is independent of $\gamma$,
these strings are interpreted as corresponding to pure gauge
theory effects.

\end{abstract}

\vspace*{.8cm}

\newpage

\sect{Introduction}

In the recent years the AdS/CFT correspondence \cite{ads/cft}
became the major analytical tool for studying gauge theories at
strong coupling. Thus it is of great interest to extend it to less
supersymmetric theories which are phenomenologically more
appropriate.

Recently Lunin and Maldacena \cite{lunmal} found the supergravity
dual of the $\b$-deformed $\N=4$ SYM \cite{leistra} by performing
an $SL(3,R)$ transformation on the well known $\axs$ background.
This provides new possibilities for quantitative checks of the
AdS/CFT correspondence. When the deformation parameter $\b$ is
real (these are called $\g$ deformations) we have a continuous
parameter which controls the deformation. So we can extend our
knowledge for the correspondence between $\axs$ and $\N=4$ SYM to
this deformed case. This idea was first explored in \cite{froits}
where the string energies in the $su(2)$ sector where matched to
the anomalous dimension of the corresponding operator in the
$\g$-deformed $\N=4$ SYM (This technique was applied to $\axs$
\cite{ts}-\cite{rashhri1}, similar considerations for other less
supersymmetric backgrounds can be found in
\cite{rash1}-\cite{pope}). In \cite{frolov} the integrability of
the bosonic string on the Lunin-Maldacena background was proven
and a new deformed background with three real deformations
parameters $\g_i$ was found. The $su(3)$ sector of semiclassical
strings in this more general three parameter background was
studied in \cite{tseytfrroi} and a remarkable match between string
energies and anomalous dimensions was shown. Different
semiclassical rotating strings with spins in both the $AdS_5$ and
the $\g$-deformed $S^5$ were studied in \cite{bodira}. More
interesting results in the study of this deformed theories can be
found in \cite{pp-wave}-\cite{many2}.

Following the ideas in \cite{lunmal} Gursoy and Nunez \cite{nunez}
found a new interesting background which is a $\b$-deformation of
the well known Maldacena-Nunez background \cite{MN}. The
Maldacena-Nunez background is dual to $N=1$ SYM and the authors of
\cite{nunez} provided evidence that the new deformed background is
dual to the so called "dipole deformation" of the field theory.
Moreover they were able to show that in the case of real
deformation the pure gauge dynamics does not depend on the
deformation parameter and the Kaluza-Klein(KK) modes increase
their masses (respectively their energy). This is a new idea which
gives a possibility to approach the problem of the mixing between
the KK modes and the pure gauge theory effects. Motivated by this
idea\footnote{We thank Carlos Nunez for suggesting this problem to
us}. For $\beta$-deformations and KK modes see also \cite{pal}, in
the present paper we want to consider semiclassical strings in the
deformed Maldacena-Nunez background and analyze the dependence of
their energies on the deformation parameter\footnote{For extended discussion about deformations of flows from type IIB supergravity see also \cite{V-P}.}. The idea is to
analyze the role of the KK modes unavoidably present in string
backgrounds with less supersymmetry. The key point is the
conjecture suggested by Gursoy and Nunez that the gamma
deformations affect only the contributions from KK modes and hence
it can serve as a test for whether our results are purely gauge
theory effects or not. The conjecture explores the following
observations. Starting from particular string background, in order
to use AdS/CFT correspondence one should consider a UV completion
of the corresponding supergravity solution. The completion, when
we consider N=1 case, depends on the KK modes. Deforming the
background we  change the behavior of the string theory (i.e. we
are producing a family of completions parameterized by a
continuous parameter $\gamma$) which results in changes in the
spectrum, the functional relations between the spins $J_1, J_2$
and the energy etc. To make definite conclusions about gauge
theory we should ensure that the result is entirely due to gauge
theory effects. The satement made in \cite{nunez} is that the
deformation affect {\it only} the dynamics of the KK modes.
Therefore, from practical point of view, one can trust only the
result which do {\it not} depend on the deformation parameter,
i.e. one should make computations in the undeformed and deformed
backgrounds and the results depending on $\gamma$ have
contributions from the KK modes, but the others are purely gauge
theory effects.

In this paper we consider the deformed Maldacena-Nunez background
and study the impact of the deformations on the semiclassical
rotating and pulsating strings. Semiclassical strings in the
Maldacena-Nunez background were studied in \cite{ponstal}. There
the cases of rotating and pulsating strings were considered and
the energy of the strings were expressed in terms of their angular
momenta and quantum numbers respectively. In sections 3 and 4 of
the present paper we consider the same ansatz as in \cite{ponstal}
both for rotating and pulsating strings. We find the energies of
the rotating strings in terms of their angular momenta in the
limit of long and short strings. For the pulsating string ansatz
we use the Bohr-Somerfeld quantization to compute the energies of
the strings in terms of their quantum number $n$ and the winding
number $m$. In both cases we reproduce the results of Pons and
Talavera in the limit $\g\rightarrow 0$, this should be expected
since this is the limit in which the background is not deformed
(i.e. the original Maldacena-Nunez background is reproduced). More
interestingly in both cases we find that the string energies
increase due to the deformatioan. This can be interpreted as an
evidence for the prediction of Gursoy and Nunez \cite{nunez} for
better decoupling of the KK modes in the deformed theory. In
\cite{nunez} it was conjectured that the sectors in which the
deformation is decoupled should correspond to pure gauge theory
effects. Fortunately, we were able to find a particular string
ansatz in which there is non-trivial decoupling of the gamma
deformation from the rest. In the concluding section we give
detailed comments on the results of our study.

\sect{$\t{\g}$ transformation of the Maldacena-Nunez background}

Here we will briefly review the background obtained by applying
the Lunin-Maldacena procedure \cite{lunmal} to the so called
Maldacena-Nunez background \cite{MN}. This $\gamma$ deformed
background was found in \cite{nunez} and it was argued that it is
dual to dipole deformations of $N=1$ SYM theory. Moreover the
authors of \cite{nunez} conjectured that in this deformed
background the KK modes decouple better from the pure gauge theory
excitations and this is controlled by the real deformation
parameter\footnote{See \cite{nunez} for an extensive discussion of
this argument and some evidences supporting it}. Following the
notation of \cite{nunez} we can write the metric of the
Maldacena-Nunez background in the following form (this is
different from the form in which the original solution was written
but it is more convenient for applying the $SL(3,R)$
transformation):

\eq{\begin{array}{l}

ds^2_{string}=
e^{\f}[dx_{1,3}^2+\a'g_sNdr^2]+D_1d\psi^2+D_2d\theta^2+D_3d\t{\theta}+E_1d\theta
d\t{\theta}+\\\\
E_2d\theta
d\psi+E_3d\t{\theta}d\psi+\ds\frac{F}{\sqrt{\Delta}}[d\varphi+(\a_1-C\b_1)d\theta+(\a_2-C\b_2)d\t{\theta}+(\a_3-C\b_3)d\psi-Cd\t{\varphi}]^2+\\\\
F\sqrt{\Delta}[d\t{\varphi}+\b_1d\theta+\b_2d\t{\theta}+\b_3d\psi]^2

\end{array}\label{0.1}}

If we define \eq{
f=4e^{2h}\sin^2\theta+\cos^2\theta+a^2\sin^2\theta \qquad
g=a\sin\theta\sin\t{\theta}\cos\psi-\cos\theta\cos\t{\theta}
\label{0.2}}

We can write the functions appearing in (\ref{0.1}) as
\eq{\begin{array}{l}

F=\ds\frac{\a'g_sNe^{\f}}{4}\sqrt{f-g^2}; \qquad
\Delta=\ds\frac{f-g^2}{f^2}; \qquad C=\frac{g}{f}; \qquad
a(r)=\ds\frac{2r}{\sinh2r};\\\\
\b_1=\ds\frac{f}{f-g^2}a\sin\psi\sin\t{\theta}; \qquad
\b_2=\ds\frac{g}{f-g^2}a\sin\psi\sin\theta; \qquad
\b_3=\ds\frac{f\cos\t{\theta}+g\cos\theta}{f-g^2};\\\\
\a_1=\ds\frac{ag\sin\t{\theta}\sin{\psi}}{f-g^2}; \qquad
\a_2=\ds\frac{a\sin\theta\sin{\psi}}{f-g^2}; \qquad
\a_3=\ds\frac{\cos\theta+g\cos\t{\theta}}{f-g^2};\\\\
D_1=\ds\frac{\a'g_sNe^{\f}}{4(f-g^2)}[f\sin^2\t{\theta}-g^2-\cos^2\theta-2g\cos\theta\cos\t{\theta}];\\\\
D_2=\ds\frac{\a'g_sNe^{\f}}{4}[a^2+4e^{2h}-\ds\frac{f}{f-g^2}a^2\sin^2\psi\sin^2\t{\theta}];
\,\,\,\,
D_3=\ds\frac{\a'g_sNe^{\f}}{4}[1-\ds\frac{a^2\sin^2\theta\sin^2\psi}{f-g^2}];\\\\
E_1=\ds\frac{a\a'g_sNe^{\f}}{2}[\cos\psi-\ds\frac{g}{f-g^2}a\sin^2\psi\sin\theta\sin\t{\theta}];\\\\
E_2=-\ds\frac{a\a'g_sNe^{\f}}{2}\ds\frac{\sin\psi\sin\t{\theta}(f\cos\t{\theta}+g\cos\theta)}{f-g^2};\\\\
E_3=-\ds\frac{a\a'g_sNe^{\f}}{2}\ds\frac{\sin\psi\sin\theta(\cos\theta+g\cos\t{\theta})}{f-g^2}
\end{array}\label{0.3}}

There are other fields present in the background but we are not
going to present their explicit form. After performing an
$SL(3,R)$ transformation (which is equivalent to a TsT
transformation \cite{frolov}) the metric takes the form:
\eq{\begin{array}{l}

(ds^2_{string})'=
\left(\ds\frac{e^{2(\Phi'-\Phi)}F}{F'}\right)^{\frac{1}{3}}[
e^{\f}[dx_{1,3}^2+\a'g_sNdr^2]+D_1d\psi^2+D_2d\theta^2+D_3d\t{\theta}+E_1d\theta
d\t{\theta}+\\\\
E_2d\theta
d\psi+E_3d\t{\theta}d\psi]+\ds\frac{F'}{\sqrt{\Delta}}[d\varphi+(\a_1-C\b_1)d\theta+(\a_2-C\b_2)d\t{\theta}+(\a_3-C\b_3)d\psi-Cd\t{\varphi}]^2+\\\\
F'\sqrt{\Delta}[d\t{\varphi}+\b_1d\theta+\b_2d\t{\theta}+\b_3d\psi]^2

\end{array}\label{0.4}}
where $\Phi$ and $\Phi'$ are the original and the transformed
dilaton field, $F'=\ds\frac{F}{1+\t{\g}^2F^2}$ and $\t{\g}$ is the
real transformation parameter. We would like to note that it can
be proven that this new deformed geometry does not generates any
new singularities with respect to the undeformed one. After this
brief comments on the deformed Maldacena-Nunez background we
proceed with the investigations of the behavior of various
classical string configurations in this geometry.

\sect{Rotating strings in $S^2\times \mathbb{R}$}

In the following sections we are going to study different string
configurations in the above presented $\t{\g}$-deformed
background. We will focus our attention on the dependence of the
string energy on the real deformation parameter $\g$ since we want
to test the proposal made in \cite{nunez} for better decoupling of
the KK modes. Since we want to compare our results to the case of
the pure Maldacena-Nunez background, we will work with the same
string configurations as those studied by Pons and Talavera
\cite{ponstal}. We start with the following simple rotating string
ansatz:

\eq{ t=\k\tau \qquad \vf=\k\o\tau \qquad \th=\ds\frac{\pi}{2}
\qquad r=r(\s) \label{1.1}} The relevant part of the deformed
metric is:

\eq{ds^2=\left(\ds\frac{e^{2(\Phi'-\Phi)}F}{F'}\right)^{\frac{1}{3}}\left(-e^{\Phi}dt^2+e^{\Phi}Adr^2\right)+\ds\frac{F'}{\sqrt{\Delta}}d\vf^2
\label{1.2}} where \eq{\begin{array}{c} A=\a'g_sN \qquad
F=\ds\frac{Ae^{\Phi}}{4}\sqrt{4e^{2h}+a^2 (r)} \qquad
F'=\ds\frac{F}{1+\t{\g}^2F^2}\\\\
e^{2\Phi'}=\ds\frac{e^{2\Phi}}{1+\t{\g}^2F^2} \qquad
\Delta=\ds\frac{1}{4e^{2h}+a^2 (r)} \qquad
a(r)=\ds\frac{2r}{\sinh2r}\\\\
e^{2h}=r\coth2r-\ds\frac{r^2}{\sinh^22r}-\ds\frac{1}{4} \qquad
e^{2\Phi}=e^{2\Phi_0}\ds\frac{sinh2r}{2e^h}
\label{1.3}\end{array}} For later convenience we define
\eq{\b(r)=e^{2h(r)}+\ds\frac{a^2 (r)}{4} \label{1.4}} Using these
definitions we can rewrite the components of the metric as follows

\eq{\begin{array}{c}
G_{tt}=-e^{\Phi} \qquad\qquad G_{rr}=Ae^{\Phi}\\\\
G_{\vf\vf}=\ds\frac{Ae^{\Phi}\b(r)}{1+\ds\frac{\t{\g}^2}{4}A^2e^{2\Phi}\b(r)}
\label{1.5}\end{array}}

The Polyakov action for this ansatz is simply

\eq{ S=-\ds\frac{1}{4\pi}\int d\tau d\s \left(e^{\Phi}\k^2 +
Ae^{\Phi}(r'(\s))^2 -
\ds\frac{Ae^{\Phi}\b(r)\k^2\o^2}{1+\ds\frac{\t{\g}^2}{4}A^2e^{2\Phi}\b(r)}
\right) \label{1.6}}

We should also impose the Virasoro constraints in order to ensure
conformal invariance. One of them is trivially satisfied by our
ansatz and from the second one we obtain the following equation
for $r(\s)$:

\eq{A(r'(\s))^2=\k^2\left(1-\ds\frac{A\b(r)\o^2}{1+\ds\frac{\t{\g}^2}{4}A^2e^{2\Phi}\b(r)}\right)\label{1.7}}

In the limit $\t{\g}\rightarrow 0$ we recover the equation for the
pure Maldacena-Nunez case, as should be expected. This equation
can be rewritten as:

\eq{d\s=dr\ds\frac{\sqrt{A}}{\k}\ds\frac{1}{\sqrt{1-\ds\frac{A\b(r)\o^2}{1+\ds\frac{\t{\g}^2}{4}A^2e^{2\Phi}\b(r)}}}
\label{1.8}}

This relation is useful for the computation of the conserved
charges in our problem (the energy and the angular momentum) which
adopt the form:

\eq{\begin{array}{c}
E=\ds\frac{\k}{2\pi}\int d\s e^{\Phi}\\\\
J=\ds\frac{A\k\o}{2\pi}\int
d\s\ds\frac{e^{\Phi}\b(r)}{1+\ds\frac{\t{\g}^2}{4}A^2e^{2\Phi}\b(r)}
\label{1.9}\end{array}}

In order to compute these quantities we need the explicit form of
the dilaton field and the function $\b(r)$

\eq{\begin{array}{c}
e^{2\Phi}=\ds\frac{e^{2\Phi_0}\sinh^22r}{\sqrt{4r\cosh2r\sinh2r-\sinh^22r-4r^2}}\\\\
\b(r)=r\coth2r-\ds\frac{1}{4} \label{1.10}\end{array}} where
$\Phi_0$ is the value of the dilaton field at the point $r=0$.
Substituting these function into the expressions for the energy
and the angular momentum leads to analytically non-solvable
integrals and we should make some approximations. Since we are
interested in the way which the deformation parameter affects the
energy we will consider the limits of long and short strings and
try to extract the leading terms for the energy and angular
momentum.

\subsection{Long strings}
The limit $r_0\rightarrow\infty$, where $r_0$ is the turning point
is called \textit{long strings}. In this limit the dilaton and the
function $\b(r)$ have the following behavior:

\eq{\b(r)\approx r \qquad e^{\Phi(r)} \approx
\ds\frac{e^r}{2r^{\frac{1}{4}}} \label{1.11}}

So in this approximation the conserved charges are

\eq{\begin{array}{c}
E=\ds\frac{\sqrt{A}}{2\pi}\int^{r_0}_{0}dr\ds\frac{e^r}{2r^{\frac{1}{4}}}\ds\frac{1}{\sqrt{1-\ds\frac{16\o^2Ar}{16+\t{\g}^2A^2e^{2r}\sqrt{r}}}}\\\\
J=\ds\frac{A^{\frac{3}{2}}\o}{2\pi}\int^{r_0}_{0}dr\ds\frac{e^rr^{\frac{3}{4}}}{2(1+\ds\frac{\t{\g}^2}{16}A^2e^{2r}\sqrt{r})}\ds\frac{1}{\sqrt{1-\ds\frac{16\o^2Ar}{16+\t{\g}^2A^2e^{2r}\sqrt{r}}}}
\label{1.12}\end{array}}

Using the long string approximation we can solve this integrals,
after some calculations we end up with the following expressions:

\eq{\begin{array}{l} E\approx
\ds\frac{\sqrt{A}}{4\sqrt{\pi}}\ds\frac{\G(3/4)}{\G(5/4)}r_0^{3/4}
{}_{1}F_{1}[\frac{3}{4},\frac{5}{4},r_0]\overset{r_0\gg}\to\ds\frac{\sqrt{A}}{4\sqrt{\pi}}r_0^{1/4}e^{r_0}\\\\
\o J\approx
\ds\frac{\sqrt{A}}{4\sqrt{\pi}}r_0^{3/4}\ds\frac{\G(7/4)}{\G(9/4)}{}_{1}F_{1}[\frac{7}{4},\frac{9}{4},r_0]\overset{r_0\gg}\to\ds\frac{\sqrt{A}}{4\sqrt{\pi}}r_0^{1/4}e^{r_0}\\\\
E-\o J\approx
\ds\frac{\sqrt{A}}{8\sqrt{\pi}}r_0^{3/4}\ds\frac{\G(3/4)}{\G(9/4)}{}_{1}F_{1}[\frac{3}{4},\frac{9}{4},r_0]\overset{r_0\gg}\to\ds\frac{\sqrt{A}}{8\sqrt{\pi}}r_0^{-3/4}e^{r_0}
\end{array}\label{1.13}}

Our results look exactly the same as those of Pons and Talavera
but in our case the turning point is:

\eq{ r_0=\ds\frac{16+\t{\g}^2A^2e^{2r_0}\sqrt{r_0}}{\o^2A}
\label{1.14}} So the effect of the deformation parameter
essentially increases the energy and the angular momentum. Of
course in the limit $\t{\g}\rightarrow 0$ we reproduce exactly the
results from the undeformed case.

We can easily calculate the ratio:

\eq{ \ds\frac{E-\o J}{\o J}\approx \ds\frac{1}{2r_0} \label{1.15}}
So as in the undeformed case this is a finite value. It aslo
useful to extract the dependance of the string energy on the
angular momentum. The leading term in this dependance (we assume
large values of $E$ and $J$) is:

\eq{ E\approx
\ds\frac{1}{R}\ds\frac{J}{\ln^{1/2}\left(\ds\frac{4\sqrt{\pi}}{R}\o
J\right)}\sqrt{16+\t{\g}^2R^2 16\pi
\o^2J^2\ln^{1/2}\left(\ds\frac{4\sqrt{\pi}}{R}\o J\right)}
\label{1.16}}

One can check that this messy expression reduces to the leading
term for the energy in the work of Pons and Talavera.

\subsection{Short strings}

Here we will present a similar analysis to the previous section
but in the limit of \textit{short strings} (i.e. $r_0\rightarrow
0$). In this limit the dilaton field and the function $\b(r)$ have
the following behavior:

\eq{\b(r)\approx \ds\frac{1}{4}+\ds\frac{2}{3}r^2 \qquad
e^{\Phi(r)} \approx 1+\ds\frac{4}{9}r^2 \label{1.17}} Substituting
these approximate values into the expressions for the energy and
the angular momentum leads to:

\eq{\begin{array}{c}
E=\ds\frac{\sqrt{A}}{2\pi\sqrt{\xi}}\int^{r_0}_{0}dr\ds\frac{\left(1+\ds\frac{4}{9}r^2\right)}{\sqrt{r_0^2-r^2}}\\\\
J=\ds\frac{A^{\frac{3}{2}}\o}{2\pi\sqrt{\xi}}\int^{r_0}_{0}dr\ds\frac{\left(\ds\frac{1}{4}+\ds\frac{7}{9}r^2\right)}{\sqrt{r_0^2-r^2}}
\label{1.18}\end{array}} where $r_0$ is the turning point and
$\xi$ is a short notation for the following expression:

\eq{\begin{array}{c}
\xi=\ds\frac{32A(3\o^2-\t{\g}^2A)}{9(16+\t{\g}^2A^2)}\\\\
r_0^2=\ds\frac{9(16+\t{\g}^2A^2-4\o^2A)}{32A(3\o^2-\t{\g}^2A)}
\end{array}\label{1.19}}
After some straightforward but tedious calculations we end up with
the following expressions for the energy and angular momentum:
\eq{\begin{array}{c}
 E\approx
\ds\sqrt{\frac{3}{2}}\frac{1}{\o}\sqrt{\ds\frac{1+\ds\frac{\t{\g}^2A^2}{16}}{1-\ds\frac{\t{\g}^2A}{3\o^2}}}\\\\
J\approx
\ds\sqrt{\frac{3}{2}}\frac{1}{\o^2}\left(\ds\frac{1+\ds\frac{\t{\g}^2A^2}{16}}{1-\ds\frac{\t{\g}^2A}{3\o^2}}\right)^{\frac{3}{2}}
\end{array}\label{1.20}}
so that we can extract the following relation

\eq{
E^2=\ds\sqrt{\frac{3}{2}}\left(\ds\frac{1+\ds\frac{\t{\g}^2A^2}{16}}{1-\ds\frac{\t{\g}^2A}{3\o^2}}\right)^{-\frac{1}{2}}J
\label{1.21}}

This is an expected Regge type behavior for the energy of a very
short string, as before if we take the limit $\t{\g}\rightarrow 0$
we reproduce the analogous relation found by Pons and Talavera.
Looking at the expression for the energy (\ref{1.20}) we see that
the deformation parameter again increases the energy and the
angular momentum. Thus we can conclude that for this rotating
string configuration the energies are shifted in the positive
direction, which comes as an evidence for the conjecture for
better decoupling of the KK modes in the $\g$-deformed geometry.
Encouraged by this observation and looking for similar pattern we
proceed with the study of semiclassical pulsating strings.

\sect{Pulsating strings in $S^2\times\mathbb{R}$}

Pulsating strings represent another interesting class of
semiclassical strings used as a tool for studying the $AdS/CFT$
correspondence \cite{arts}-\cite{smedb}. We use the approach of
Minahan \cite{min} and the following pulsating string ansatz: \eq{
t=\tau \qquad \vf=m\s \qquad \th=\ds\frac{\pi}{2} \qquad r=r(\tau)
\label{2.1}} The Nambu-Goto Lagrangian for this ansatz is
\eq{\L_{NG}=me^{\Phi}\sqrt{\ds\frac{A\b(r)(1-A\dot{r}^2)}{1+\ds\frac{\t{\g}^2}{4}A^2e^{2\Phi}\b(r)}}
\label{2.2}} The canonical momentum corresponding to $r$ can be
obtained from this Lagrangian:
\eq{\Pi_r=\ds\frac{1}{1+\ds\frac{\t{\g}^2}{4}A^2e^{2\Phi}\b(r)}\ds\frac{me^{\Phi}A^2\b(r)\dot{r}}{\sqrt{\ds\frac{A\b(r)(1-A\dot{r}^2)}{1+\ds\frac{\t{\g}^2}{4}A^2e^{2\Phi}\b(r)}}}
\label{2.3}} Now we can easily compute the Hamiltonian of our
system \eq{
H=\sqrt{\ds\frac{\Pi^2_r}{A}+\ds\frac{m^2Ae^{2\Phi}\b(r)}{1+\ds\frac{\t{\g}^2}{4}A^2e^{2\Phi}\b(r)}}
\label{2.4}} This is the Hamiltonian of a one dimensional system
with the potential: \eq{
V=\ds\frac{m^2A^2e^{2\Phi}\b(r)}{1+\ds\frac{\t{\g}^2}{4}A^2e^{2\Phi}\b(r)}
\label{2.5}} In order to compute the energy levels we will make
use of the WKB approximation, namely
\eq{I=\int^{r_0}_0dr\sqrt{AE^2-\ds\frac{m^2A^2e^{2\Phi}\b(r)}{1+\ds\frac{\t{\g}^2}{4}A^2e^{2\Phi}\b(r)}}=(n+\frac{1}{2})\pi
\label{2.6}} Unfortunately this integral is analytically
unsolvable and we should again make use of some appropriate
approximations. As in the previous section we will focus on the
limits of short and long strings and find the leading
contributions to the energy.

\subsection{Long strings}
The limit of long strings $r_0\rightarrow \infty$ is physically
interesting and it greatly simplifies the form of the potential,
which has the following asymptotic behavior: \eq{
V=\ds\frac{\ds\frac{m^2A^2}{4}e^{2r}\sqrt{r}}{1+\ds\frac{\t{\g}^2A^2}{16}e^{2r}\sqrt{r}}
\label{2.7}} Now we can solve the equation $AE^2=V(r_0)$ for the
turning point $r_0$, of course we again explore the approximation
$r_0\rightarrow \infty$. After some computations we find:
\eq{r_0\approx
\ds\frac{1}{2}\ln\left(\frac{4E^2}{m^2A-\ds\frac{\t{\g}^2A^2E^2}{4}}\right)\label{2.8}}
The integral $I$ in this limit has the following leading behavior:
\eq{ I\approx \sqrt{A}Er_0+\O(\ds\frac{1}{E}) \label{2.9}} So we
can substitute our expression for $r_0$ and end up with:
\eq{I\approx
\ds\frac{1}{2}\sqrt{A}E\ln\left(\frac{4E^2}{m^2A-\ds\frac{\t{\g}^2A^2E^2}{4}}\right)+\O(\ds\frac{1}{E})\label{2.10}}
We can compare our result to the one from the paper of Pons and
Talavera. In the limit $\t{\g}\rightarrow 0$ we reproduce their
result as expected, because this is the limit which reproduces the
original Maldacena-Nunez background. Moreover since
$I=(n+\frac{1}{2})\pi$ we see that the energy levels are again
shifted upwards, which coincides with the pattern found in the
previous section for rotating strings.

\subsection{Short strings}
We will consider the limit $r_0\rightarrow 0$ (i.e. \textit{short
strings}). In this limit the potential adopts the following
simpler form:
\eq{V(r)=m^2A^2\left(\ds\frac{1}{1+\ds\frac{\t{\g}^2A^2}{4}}+r^2\right)
\label{2.11}} Using this we can find the turning point $r_0$, i.e.
the solution of the equation $E^2=V(r_0)$
\eq{r_0=\ds\frac{\sqrt{A(4+\t{\g}^2A^2)(4E^2+E^2\t{\g}^2A^2-4m^2A)}}{m(4A+\t{\g}^2A^3)}
\label{2.12}} Now the integral $I=\int^{r_0}_0dr\sqrt{E^2-V(r)}$
can be solved but the result is to long and we will not present it
here. The important thing is that we can further solve the
equation for $E^2$ in terms of the oscillator level $n$ \eq{E^2
\approx
\ds\left(\frac{\sqrt{A}+A^{\frac{5}{2}}\t{\g}^2}{1+\ds\frac{\t{\g}^2A^2}{4}}\right)mn
\label{2.13}} Amazingly in the limit $\t{\g}\rightarrow 0$ this
expression reduces to the result of Pons and Talavera. However
here it is not that obvious that the energy is increased but since
in the undeformed case $E^2=\sqrt{A}mn$ we can say that
(\ref{2.13}) is affected by the deformation parameter and indeed
it is greater than the energy of the string in the undeformed
geometry.

\sect{" Fast " rotating strings in $S^3$}

In this case, the discussion follows closely the one in the
$\gamma$-deformed $S^3$ part. If we put
$$\theta\,=\,0,\,\,r\,=\,0,\,\, \varphi\,=\,const$$
the relevant part of the metric is
\begin{equation}
ds^2=-e^{\Phi_0}dt^2+Ad\tilde{\theta}^2+\frac{A}{1+\gamma^2A^2\sin^2\tilde{\theta}}\left(d\psi+\cos\tilde{\theta}d\tilde{\varphi}\right)^2+\frac{A\sin^2\tilde{\theta}}{1+\gamma^2A^2\sin^2\tilde{\theta}}\,d\tilde{\varphi}^2,\label{F1}
\end{equation}
where $A=\frac{\alpha^{\prime}g_s\,N\,e^{\Phi_0}}{4}$.\\
The B-field is $B_2\,=\,d\psi\,\wedge\,d\tilde{\varphi}$.\\
Let us now study the structure of the string sigma model action in
the "2-spin" sector in the limit of large total spin $J=J_1+J_2$.
Setting \eq{\psi=\alpha+\xi,\qquad \tilde{\varphi}=\alpha-\xi,\,\,
i.\,e.
\,\,\,\,\xi\equiv\frac{1}{2}(\psi-\tilde{\varphi}),\label{F2}}
 we may
treat $\alpha$ as a "fast" angular variable while $\xi$ and
$\tilde{\theta}$ will be "slow" variables whose time evolution
will be suppressed. Using these definitions we can rewrite the
metric as follows
\begin{equation}
ds^2=-e^{\Phi_0}dt^2+Ad\tilde{\theta}^2+\ds\frac{4A}{1+\gamma^2A^2\sin^2\tilde{\theta}}\left[\cos^2(\ds\frac{\tilde{\theta}}{2})d\alpha^2+\sin^2(\ds\frac{\tilde{\theta}}{2})d\xi^2\right].\label{F3}
\end{equation}
The B-field is
$B_2=d\psi\,\wedge\,d\tilde{\varphi}=-2\,d\alpha\,\wedge\,d\xi$,
i. e.
$B_{\alpha\xi}=-2$.\\
Then the relevant Polyakov action is
\begin{multline}
S=-\ds\frac{1}{4\pi}\int d\tau
d\sigma\,[\,\sqrt{-h}h^{pq}\,(\,-e^{\Phi_0}\partial_pt\partial_qt+A\,\partial_p\tilde{\theta}\partial_q\tilde{\theta}+\\+\ds\frac{4A\cos^2(\ds\frac{\tilde{\theta}}{2})}{1+\gamma^2A^2\sin^2\tilde{\theta}}\,\partial_p\alpha\partial_q\alpha+
\ds\frac{4A\sin^2(\ds\frac{\tilde{\theta}}{2})}{1+\gamma^2A^2\sin^2\tilde{\theta}}\,\partial_p\xi\partial_q\xi\,)-2\varepsilon^{pq}B_{\alpha\xi}\,\partial_p\alpha\partial_q\xi\,].\label{F4}
\end{multline}
To implement the uniform gauge fixing one may either consider the
phase-space action and fix $p_\alpha=const$ or, equivalently,
first do T-duality (i.e. 2-d duality) in $\alpha$ direction in the
above action and gauge-fix $\tilde{\alpha}=J\sigma$. The
components of the metric \eqref{F3} after T-duality transformation
take the form: \eq{\begin{array}{c}
\tilde{G}_{\tilde{\alpha}\tilde{\alpha}}\,=\,\ds\frac{1+\gamma^2A^2\sin^2\tilde{\theta}}{4A\cos^2(\ds\frac{\tilde{\theta}}{2})}\,\equiv\,g_1,\qquad
\tilde{G}_{tt}\,=\,-e^{\Phi_0},\qquad
\tilde{G}_{\tilde{\theta}\tilde{\theta}}\,=\,A,\\\\
\tilde{G}_{\xi\xi}=\ds\frac{A^2\sin^2\tilde{\theta}+(1+\gamma^2A^2\sin^2\tilde{\theta})^2}{A\cos^2(\ds\frac{\tilde{\theta}}{2})\,(1+\gamma^2A^2\sin^2\tilde{\theta})}\,\equiv\,g_2,\,\,
\tilde{G}_{\tilde{\alpha}\xi}=-2\ds\frac{(1+\gamma^2A^2\sin^2\tilde{\theta})}{4A\cos^2(\ds\frac{\tilde{\theta}}{2})}=-2g_1
\label{F5}\end{array}} and then we have
\eq{d{\tilde{s}}^2=-e^{\Phi_0}\,dt^2+A\,d\tilde{\theta}^2+g_1\,d{\tilde{\alpha}}^2+g_2\,d\xi^2-4g_1\,d{\tilde{\alpha}}d\xi.\label{F6}}
All the components of $\tilde{B}$-field are zero.

The action after T-duality takes the form:
 \ml{S=-\ds\frac{1}{4\pi}\int d\tau
d\sigma\,\sqrt{-h}h^{pq}\,[\,-e^{\Phi_0}\partial_pt\partial_qt+A\,\partial_p\tilde{\theta}\partial_q\tilde{\theta}+g_1\,\partial_p\tilde{\alpha}\partial_q\tilde{\alpha}+\\+g_2\,\partial_p\xi\partial_q\xi-2g_1\,\partial_p\tilde{\alpha}\partial_q\xi\,].\label{F7}}
We are going to use Nambu-Goto action, because we don't know the
gauge-fixing of the $\sqrt{-h}h^{pq}$.\\
Imposing now the gauge ( here $\tau$ and $\sigma$ are world-sheet
coordinates ) \eq{t=\tau,\qquad \tilde{\alpha}=J\sigma,\label{F8}}
and solving for the world-sheet metric, we find
\ml{-det\,h_{pq}=\,[\,A\,\dot{\tilde{\theta}}{\tilde{\theta}}^{\prime}+g_2\dot{\xi}{\xi}^{\prime}-2g_1J\dot{\xi}\,]^2\\
-[\,-e^{\Phi_0}+A\dot{{\tilde{\theta}}^2}+g_2{\dot{\xi}}^2]\,[A{\tilde{\theta}}^{{\prime}^2}+g_1J^2+g_2{\xi}^{{\prime}^2}-4g_1J{\xi}^{\prime}\,].\label{F9}}
In the simple case $\tilde{\theta}= const$ we have
\eq{\sqrt{-det\,h_{pq}}=\sqrt{e^{\Phi_0}g_1J^2-(g_1\,g_2-4g_1^2)\,J^2{\dot{\xi}}^2+e^{\Phi_0}g_2{\xi}^{{\prime}^2}-4e^{\Phi_0}g_1J\xi^\prime}.\label{F10}}
To isolate the sector of "fast strings" we should take J to be
large, i.e. $\frac{1}{J^2}\,\rightarrow\,0$, and $\dot{\xi}$ to be
small ($\xi$ is "slow" variable). Then

\ml{\sqrt{-det\,h_{pq}}=J\,\sqrt{e^{\Phi_0}g_1-(g_1\,g_2-4g_1^2)\,{\dot{\xi}}^2-4e^{\Phi_0}g_1\xi^\prime\ds\frac{1}{J}+e^{\Phi_0}g_2{\xi}^{{\prime}^2}\ds\frac{1}{J^2}}=\\
=J\,\left[\sqrt{e^{\Phi_0}g_1-(g_1\,g_2-4g_1^2)\,{\dot{\xi}}^2}-\ds\frac{4e^{\Phi_0}g_1\xi^\prime}{2\sqrt{e^{\Phi_0}g_1-(g_1\,g_2-4g_1^2)\,{\dot{\xi}}^2}}\,\ds\frac{1}{J}+O(\ds\frac{1}{J})\right]\approx\\
\approx\left[\,J\,\left(e^{\Phi_0}g_1-(g_1\,g_2-4g_1^2)\,{\dot{\xi}}^2\right)-2e^{\Phi_0}g_1\xi^\prime\right]\,\ds\frac{1}{\sqrt{e^{\Phi_0}g_1-(g_1\,g_2-4g_1^2)\,{\dot{\xi}}^2}}=\\
=\left[\,J\,\left(e^{\Phi_0}g_1-(g_1\,g_2-4g_1^2)\,{\dot{\xi}}^2\right)-2e^{\Phi_0}g_1\xi^\prime\right]\left[\ds\frac{1}{e^{\Phi_0/2}\sqrt{g_1}}+\ds\frac{(g_1\,g_2-4g_1^2)}{2(e^{\Phi_0}g_1)^{3/2}}\,{\dot{\xi}}^2+o({\dot{\xi}}^4)\right]\\\\
\approx\,-\ds\frac{\sqrt{g_1}(g_2-4g_1)}{2\,e^{{\Phi_0/2}}}\left(J+2\xi^\prime\right)\,{\dot{\xi}}^2-2\,e^{{\Phi_0/2}}\,\sqrt{g_1}\,\xi^\prime+J\,e^{{\Phi_0/2}}\,\sqrt{g_1}.\label{F11}}
Finally, the Nambu-Goto Lagrangian takes the form
\eq{\mathcal{L}({\dot{\xi}},\xi^\prime)=-\ds\frac{1}{2\pi\alpha'}\left[-\ds\frac{\sqrt{g_1}(g_2-4g_1)}{2\,e^{{\Phi_0/2}}}\left(J+2\xi^\prime\right)\,{\dot{\xi}}^2-2\,e^{{\Phi_0/2}}\,\sqrt{g_1}\,\xi^\prime+J\,e^{{\Phi_0/2}}\,\sqrt{g_1}\right].\label{F12}}
The ansatz we employ is \eq{\tilde{\theta}=const,\qquad
\xi\equiv\ds\frac{1}{2}\,(\psi-\tilde{\varphi})=\,\omega\tau+\ds\frac{1}{2}m\sigma,\label{F13}}
where $m=m_1-m_2$ is an integer and $\omega$ is a constant ( one
can check that these are solutions of the equations of motion ).

Now it is straightforward to compute the canonical momentum
corresponding to $\xi$
\eq{\Pi_{\xi}\equiv\,J_1-J_2=\ds\frac{\partial\mathcal{L}({\dot{\xi}},\xi^\prime)}{\partial\dot{\xi}}=\ds\frac{1}{2\pi\alpha'}\,\ds\frac{\sqrt{g_1}(g_2-4g_1)}{e^{{\Phi_0/2}}}\,\left(J+m\right)\,\omega\label{F14}}
Computing the energy we find
\eq{E=\Pi_{\xi}\,\dot{\xi}-\mathcal{L}({\dot{\xi}},\xi^\prime)=\ds\frac{1}{2\pi\alpha'}\left[\ds\frac{\sqrt{g_1}(g_2-4g_1)}{2e^{{\Phi_0/2}}}\,\left(J+m\right)\,\omega^2+e^{{\Phi_0/2}}\,\sqrt{g_1}\,\left(J-m\right)\right],\label{F15}}
or
\eq{E=\frac{1}{2}\,(J_1-J_2)\,\omega+\ds\frac{1}{2\pi\alpha'}\,e^{{\Phi_0/2}}\,\sqrt{g_1}\,\left(J-m\right),\label{F16}}
where
\eq{\sqrt{g_1}(g_2-4g_1)=\frac{2\sqrt{A}\sin^2({\tilde{\theta}}/2)}{\cos({\tilde{\theta}}/2)\sqrt{1+\gamma^2A^2\sin^2\tilde{\theta}}},\label{F17}}

\eq{\sqrt{g_1}=\frac{\sqrt{1+\gamma^2A^2\sin^2\tilde{\theta}}}{2\sqrt{A}\cos({\tilde{\theta}}/2)},
\qquad
g_2-4g_1=\ds\frac{4A\sin^2(\ds\frac{\tilde{\theta}}{2})}{1+\gamma^2A^2\sin^2\tilde{\theta}}\label{F18}}

\sect{Rotating strings with spins in the both cycles, $S^2$ and
$S^3$}

In this section, the discussion follows closely the one in the
$\gamma$-deformed $S^5$ part. If we put
$$\psi\,=\,0,\qquad r\,=\,0,$$
the relevant parts of the metric and the B-field are \ml{
ds^2=-e^{\Phi_0}dt^2+A\,d\,(\tilde{\theta}+\theta)^2+\\
+\frac{A}{1+\gamma^2A^2\sin^2(\tilde{\theta}+\theta)}\left(d\varphi+\cos(\tilde{\theta}+\theta)\,d\tilde{\varphi}\right)^2+
\frac{A\sin^2(\tilde{\theta}+\theta)}{1+\gamma^2A^2\sin^2(\tilde{\theta}+\theta)}\,d\tilde{\varphi}^2,\label{R1}
} \eq{B_2\,=\,d\varphi\,\wedge\,d\tilde{\varphi},\label{R2}} where
$\,A=\frac{\alpha^{\prime}g_s\,N\,e^{\Phi_0}}{4}\,$.

The ansatz \eq{t=\tau; \qquad \theta\,=\,-\,\tilde{\theta};\qquad
\varphi=\,-\,m\,\sigma+\omega_1\,\tau;\qquad
\tilde{\varphi}=\,m\,\sigma+\omega_2\,\tau,\label{R3}}
 is a solution of the equations of motion ($m$ is an integer).

We should also impose the Virasoro constraints in order to ensure
conformal invariance. One of them is trivially satisfied by our
ansatz and from the second one we obtain the following relation
for $\,\,\omega_1+\omega_2\,\equiv\,\omega$:
\eq{\omega^2=\ds\frac{e^{\Phi_0}}{A},\qquad i.e.\qquad
\omega=\ds\frac{2}{\sqrt{{\alpha}^{\prime}\,g_s\,N}}.\label{R4}}

The relevant bosonic part of the classical string action depends
on the metric $\,G_{MN}\,$ and the NS-NS 2-form B-field
$\,B_{MN},\,$ i.e. is given by
\eq{S=\,-\ds\frac{1}{4\pi}\,\int\,d\tau\,d\sigma\,\left[g^{p\,q}\,G_{MN}\,\partial_p\,X^M\partial_q\,X^N-\epsilon^{p\,q}\,B_{MN}\,\partial_p\,X^M\partial_q\,X^N\right],\label{R5}}
where $\,\epsilon^{0\,1}=1\,$ and
$\,g^{p\,q}\equiv\sqrt{-h}\,h^{p\,q}\,$($\,h^{p\,q}\,$ is a
world-sheet metric with Minkowski signature, i.e. in the conformal
gauge $\,g^{p\,q}\,=\,diag(-1,1)\,$).

The Polyakov action for this ansatz is simply
\eq{S=\,-\ds\frac{1}{4\pi}\,\int\,d\tau\,d\sigma\,\left[e^{\Phi_0}\,{\dot{t}}^{\,2}-A\,\left(\dot{\varphi}+\dot{\tilde{\varphi}}\right)^{\,2}-2\,m\,\left(\dot{\varphi}+\dot{\tilde{\varphi}}\right)\,\right].\label{R6}}

Now it is straightforward to compute the angular momenta $\,J_1\,$
and $\,J_2\,$ corresponding to $\,\varphi\,$ and
$\,\tilde{\varphi}\,$
\eq{J_1=J_2\,=\,A\,(\omega_1+\omega_2)+m=A\,\omega+m\,.\label{R7}}
Computing the energy we find \eq{E\,=\,e^{\Phi_0}\,.\label{R8}}
Using the relation \eqref{R4} and set
$\,A=\frac{\alpha^{\prime}g_s\,N\,e^{\Phi_0}}{4}\,$ we can rewrite
the energy with the total spin $\,J=J_1+J_2=2A\omega+2m\,$ as
follow
\eq{E\,=\,\ds\frac{\left(J-2m\right)\,\omega}{2}\,.\label{R9}}

We should note also that the metrics \eqref{F1} and \eqref{R1} are
identical. Changing $\,\tilde{\theta}+\theta\,$ by
$\,\tilde{\theta}\,$ and $\,\varphi\,$ by $\,\psi\,$ in metric
\eqref{R1} we obtain the metric \eqref{F1}. Consequently, the
ansatz \eq{t=\tau; \qquad \tilde{\theta}=\,0;\qquad
\psi=\,-\,m\,\sigma+\omega_1\,\tau;\qquad
\tilde{\varphi}=\,m\,\sigma+\omega_2\,\tau,\label{R10}} is a
solution of the equations of motion for the string-frame
\eqref{F1}. Therefore, the result for energy will be the same as
in \eqref{R9}.

\sect{More rotating strings}

Let us consider the following limiting case of the geometry
considered in \cite{nunez}. If in background \eqref{0.1} we take
the limit $\theta,\,\tilde{\theta},\,\varphi$ and
 $\tilde{\varphi}\,\longrightarrow\,0\,\,$ the only contributions will
come from the first and second terms in the square brackets. The
resulting geometry then becomes:
\begin{equation}
ds^2_{limit}\,=\,e^{\Phi(r)}\,\left[-dt^2+dX_1^2+dX_2^2+dZ^2+\alpha^{\prime}g_{s}N\,dr^2+\frac{\alpha^{\prime}g_{s}N}{4}\,d\psi^2\right].\label{L1}
\end{equation}

The classical equations of motion for the string sigma model are:
\al{
& \p_\a\left( e^{\Phi(r)}\eta^{\a\b}\p_\b t\right) =0 \label{eq1} \\
& \p_\a\left( e^{\Phi(r)}\eta^{\a\b}\p_\b X^i\right) =0 \label{eq2} \\
& \p_\a\left( e^{\Phi(r)}\eta^{\a\b}\p_\b r\right) = 
\frac{1}{2}\left( e^{\Phi(r)}\right)\eta^{\a\b} \left[ -\p_\a t\p_\b t+
\p_a X^i\p_b X^i\right] 
\label{eq3}
}
We make the following ansatz:
\al{
&t=\kappa\tau \notag \\
& X_i=x_i(\tau)\rho_i(\sigma) \label{eq4} \\
& r=r(\sigma) \notag
}
The equations of motion for the $X_i$ variables (\ref{eq1},\ref{eq2},\ref{eq3}) then become
\eq{
-e^{\Phi(r)}\ddot x_i(\tau) + \rho_i(\sigma)+x_i(\tau)\p_\sigma\left( e^{\Phi(r)}\right) =0 \label{eq5}
}
We can now impose natural separation of variables which leads to
\al{
&\ddot x_i(\tau)+\omega_1^2 x_i(\tau)=0 \label{eq6} \\
& \p_\sigma\left( e^{\Phi(r)} rho'(\sigma)\right) +\omega_1^2
e^{\Phi(r)}\rho'(\sigma)=0 .
\label{eq7}
}
The solutions of (\ref{eq6}) are correspondingly
\eq{
x_1=\cos\omega\tau , \qquad x_2=\sin\omega\tau
}
For the radial variable $r$ we get
\al{
\p_\sigma\left( e^{\Phi(r)}\p_\sigma r\right) = 
\frac{1}{2}\p_r e^{\Phi(r)}
\left[ \kappa^2- \right. &
\omega_1^2\rho_1^2\sin^2\omega\tau-\omega_1^2\rho_2^2\cos^2\omega\tau 
 \notag \\
& \left. +{\rho_1'}^2\cos^2\omega\tau+{\rho_2'}^2\sin^2\omega\tau\right] 
\label{eq8}
}
The  Virasoro constraints can be written in the form
\al{
& e^{\Phi(r)}{r'}^2+e^{\Phi(r)}\left[ -\kappa^2+\omega_1^2\rho_1^2\sin^2\omega_1\tau
\right. & \notag \\
& \left. \omega_1^2\rho_2^2\cos^2\omega\tau+{\rho_1'}^2\cos^2\omega\tau+ {\rho_2'}^2\sin^2\omega\tau\right] =0.
\label{eq9}
}

One can write now the corresponding conserved charges
\eq{
E=\frac{\kappa}{2\pi\a'}\int e^{\Phi(r)}d\sigma \label{eq10}
}
\al{
J= & \frac{1}{2\pi\a'}\int e^{\Phi(r)}\rho_1\rho_2\left[ 
x_1\p_\tau x_2- x_2\p_\tau x_1\right] d\sigma \notag \\
& = \frac{\omega}{2\pi\a'}\int e^{\Phi(r)}\rho_1\rho_2\,d\sigma
\label{11}
}
To solve the equations of motion we impose the following simplifying condition
\eq{
\rho_1=\rho_2 = \rho \notag
}
after which the equations of motion become
\eq{
\p_\sigma\left( e^{\Phi(r)}\p_\sigma r\right) - 
\frac{1}{2}\p_r e^{\Phi(r)}
\left[ \kappa^2- \omega_1^2\rho^2 + {\rho'}^2\right] 
\label{eq12}
}
The Virasoro constraints take the form
\eq{
e^{\Phi(r)}{r'}^2+e^{\Phi(r)}\left[ -\kappa^2+\omega_1^2\rho^2+{\rho'}^2\right] =0.
}
which gives for the angular momentum the expression
\eq{
J=\frac{\omega}{2\pi\a'}\int e^{\Phi(r)}\rho^2\,d\sigma.
}
The AdS/CFT duality teach us that the corresponding state in the gauge theory has the same quantum numbers. In order the semiclassical approximation to be valid the action should be large and thus we are dealing with a state in the IR region of the corresponding gauge theory with large momentum $J$ and energy $E$. One can consider now the string at the stationary point $r_0=0$. The equation of motion for $r$ then is trivially satisfied and the Virasoro constraints gives an equation for $\rho$
\eq{
{\rho'}^2+\omega_1^2\rho^2=\kappa^2\label{13}.
}
The solution is
\eq{
\rho(\sigma)=A\sin\omega\sigma,
}
and using again the Virasoro constraint we get for $A$
\eq{
A=\frac{\kappa}{\omega} \notag.
}
The expression for the energy and angular momentum then become
\eq{
E=2\pi\kappa T_s; \quad J=\frac{\kappa^2}{\omega_1^2}\pi T_s,
\label{14}
}
where $T_s$ is the string tension. It is obvious that we reproduce the well known Regge relation
\eq{
J=\frac{1}{4\pi T_{s,eff}}E^2.
}
One can study now the fluctuations around the given configuration and find the quantum corrections to the relation we obtained, but this is subject to another investigation. The main lesson we can extract is that having obtained a solution independent of $\gamma$, we reproduce the well known Regge behavior and one can conclude the our solution is physically reasonable. Although our result do not give conclusive solution of the decoupling of the Kaluza-Klein modes, it gives more arguments to the conjecture that $\gamma$ deformation lifts the KK-modes and one can serve as a mechasnism of separating the physical solutions.

As another test of the conjecture we consider strings with $r$ nontrivially depending on the worldsheet coordinate $\sigma$, i.e. the string is extended in $r$ direction too.

The Virasoro constraint and the equation of motion for
$\,\rho(\sigma)\,$ are:
\begin{equation}
A\,{r}^{\,\prime\,2}(\tilde{\sigma})+{\rho}^{\,\prime\,2}(\tilde{\sigma})+\rho^2(\tilde{\sigma})\,=\,\frac{B^2}{\omega_1^2},\label{H1}
\end{equation}
\begin{equation}
{\rho}^{\,\prime\,\prime}(\tilde{\sigma})+\frac{d\Phi(r)}{dr}\,{r}^{\,\prime}(\tilde{\sigma})\,{\rho}^{\,\prime}(\tilde{\sigma})+\rho(\tilde{\sigma})\,=\,0,\label{H2}
\end{equation}
where $\,\tilde{\sigma}\,=\,\omega_1\,\sigma\,$ and
$\,A\,=\,\alpha^{\prime}g_{s}N,\,\,\,\,B^2\,=\,\kappa^2-\frac{\alpha^{\prime}g_{s}N}{4}\,\omega_2^2.$

For small $\,\rho\,$ we have:
\begin{equation}
A\,{r}^{\,\prime\,2}(\tilde{\sigma})+{\rho}^{\,\prime\,2}(\tilde{\sigma})\,=\,\frac{B^2}{\omega_1^2},\label{H3}
\end{equation}
\begin{equation}
{\rho}^{\,\prime\,\prime}(\tilde{\sigma})+\frac{d\Phi(r)}{d\tilde{\sigma}}\,{\rho}^{\,\prime}(\tilde{\sigma})\,=\,0.\label{H4}
\end{equation}
Integrating equation \eqref{H4} once we obtain:
\begin{equation}
|\,{\rho}^{\,\prime}(\tilde{\sigma})\,|\,=\,C\,e^{-\Phi(r)}\label{H5}
\end{equation}
For $\,r\,$ near by to fixed $\,r_0\,$ we have the following
approximation
\begin{equation}
{\rho}^{\,\prime}(\tilde{\sigma})\,=\,-\,C\,e^{-\Phi(r_0)}\left[1-\frac{d\Phi(r_0)}{dr}\,(r-r_0)\right].\label{H6}
\end{equation}
Using the Virasoro constraint \eqref{H3} we obtain for
$\,r(\tilde{\sigma})\,$ the equation:
\begin{equation}
A\,{r}^{\,\prime\,2}(\tilde{\sigma})\,=\,\frac{B^2}{\omega_1^2}\left[1-\frac{C^2\,\omega_1^2}{B^2}\,e^{-2\Phi(r_0)}\left(1-\frac{d\Phi(r_0)}{dr}\,(r-r_0)\right)^2\right],\label{H7}
\end{equation}
when $\,r_1^\star\,\leq\,r\,\leq\,r_2^\star,\,\,$ where
\begin{align}
r_1^\star\,=\,r_0-\frac{\frac{B\,e^{\Phi(r_0)}}{C\,\omega_1}-1}{\frac{d\Phi(r_0)}{dr}},\qquad
r_2^\star\,=\,r_0+\frac{\frac{B\,e^{\Phi(r_0)}}{C\,\omega_1}+1}{\frac{d\Phi(r_0)}{dr}},\label{H8}
\end{align}
are the turning points of string and
$\,r_0+\frac{1}{\frac{d\Phi(r_0)}{dr}}\,$ is the center of mass.
This equation can be rewritten as:
\begin{equation}
\frac{d\,r}{\sqrt{1-\frac{C^2\,\omega_1^2}{B^2}\,e^{-2\Phi(r_0)}\left(1-\frac{d\Phi(r_0)}{dr}\,(r-r_0)\right)^2}}\,=\,\frac{B}{\omega_1\,\sqrt{A}}\,\,d\,\tilde{\sigma}.\label{H9}
\end{equation}
This relation is appropriate for the computation of the conserved
charges in our problem (the energy and angular momenta ).

The $2\pi$ periodic solutions of \eqref{H7} and \eqref{H6} are:
\begin{equation}
r(\sigma)\,=\,r_0+\frac{1}{\frac{d\Phi(r_0)}{dr}}\,\left[1+\frac{B\,e^{\Phi(r_0)}}{C\,\omega_1}\,\sin(m\,\sigma)\right],\label{H10}
\end{equation}
\begin{equation}
\rho(\sigma)\,=\,\frac{B\,\sqrt{A}\,e^{\Phi(r_0)}}{C\,\omega_1}\,\frac{1}{\frac{d\Phi(r_0)}{dr}}\,\left[1-\cos(m\,\sigma)\right],\label{H11}
\end{equation}
where
$\,m\,=\,\frac{C}{\sqrt{A}}\,e^{-\Phi(r_0)}\,\frac{d\Phi(r_0)}{dr}\,\omega_1\,=\,4\,k+1,\,\,\,\,k=0,1,2,3,...\,$.

For energy and spins in this case we have:
\begin{equation}
E\,=\,2\kappa\,\frac{\sqrt{A}\,e^{2\Phi(r_0)}}{C\,\frac{d\Phi(r_0)}{dr}}\,\frac{1}{\omega_1}\,\left[1+\frac{B\,e^{\Phi(r_0)}}{\pi\,C}\,\frac{1}{\omega_1}\right],\label{H12}
\end{equation}
\begin{equation}
J_2\,=\,\frac{A^{3/2}\,e^{2\Phi(r_0)}}{2\,C\,\frac{d\Phi(r_0)}{dr}}\,\frac{\omega_2}{\omega_1}\,\left[1+\frac{B\,e^{\Phi(r_0)}}{\pi\,C}\,\frac{1}{\omega_1}\right],\label{H13}
\end{equation}
\begin{equation}
J_1\,=\,\frac{2\,A\,B^2\,e^{3\Phi(r_0)}}{\pi\,C^2\,{(\frac{d\Phi(r_0)}{dr})}^2}\,\frac{1}{\omega_1}\,\left[\frac{3\pi}{2}-4\,\frac{\sqrt{A}\,e^{\Phi(r_0)}}{C\,\frac{d\Phi(r_0)}{dr}}\,\frac{1}{\omega_1}+\frac{1}{3}\,\frac{B\,\sqrt{A}\,e^{2\Phi(r_0)}}{C^2\,
\frac{d\Phi(r_0)}{dr}}\,\frac{1}{\omega_1^2}\right].\label{H14}
\end{equation}

As a result of our considerations one can conclude that there exsist indeed physical solutions that give the conserved quantities independent of the deformation parameter. Once again, we point out that these computations show that the conjecture can serve as a mechanism of separating physically relevant solutions. It is important however to point out that sometimes the background can "cheate" and one can have a realization of some "secrete" mechanism of cancelation of the contributions coming from gamma deformation with KK modes still in the game and the conjecture we checked in this paper cannot be considered as a definite way of separating pure gauge theory effects. We will discusse more on this issue in the next Section.

\sect{Conclusions}

In this paper we considered string theory in the deformed
Maldacena -Nunez background studied in \cite{nunez} at
supergravity level. We studied various string configurations of
string rotating in several cycles of the geometry and pulsating
strings as well, and discussed the effect of the deformation.

After listing the basic formulae for the deformed Maldacena-Nunez
background in Section 2, in Section 3 we present the analysis of
the simple rotating strings in the $S^2\times \mathbb{R}$. Since
the calculations cannot be given in analytic form, we consider the
limit of short and long strings. In this approximation we show
that the energy increases due to the deformation parameter
$\gamma$. We arrived to the same conclusions in Section 4 where we
consider short and long pulsating strings in the same cycle of the
geometry. Since the geometry have one more cycle in its five
dimensional part, we consider fast moving strings in $S^3\times
\mathbb{R}$. Detailed consideration show that again we have
increasing of the energy due to the deformation.

In \cite{nunez} it was conjectured that if one can separate a
sector in which the gamma deformation completely decouples, one
can state that the effect should be purely due to the gauge theory
with no contributions from KK modes. In the last section we were
able to find a nontrivial string configurations in which this can
be achieved.

This check is quite remarkable, so let us comment in some details
what we found. As we mentioned above, the conjecture made in
\cite{nunez} is that the gamma-deformed and the undeformed
background only differ in the dynamics of the KK modes. To check
the conjecture we considered various rotating and pulsating string
configurations. The first type of configurations is when strings
rotate in any combinations of $\varphi$ and $\tilde\varphi$
directions. By means of AdS/CFT correspondence these should
describe long operators in the gauge theory of the form
\eq{
\mathcal A_{KK}=\O\O\cdots\O{\mathcal K}{\mathcal K}{\mathcal K}\O\cdots\O,
\label{opers1} }
where $\O$ represents any $\mathcal N=1$ SYM
operator and $\mathcal K$ - any operator made out of KK modes. The
second type configurations is when strings do {\it not} rotate in
those directions. The statement of the conjecture is that in this
case the long operators are made out of operators of the SYM
fields
\eq{
\mathcal A=\O\O\cdots\O\O\cdots\O.
\label{opers2} }
What are the reasons for the above comments? First of all, let us mention the
dual Lagrangian of the Maldacena-Nunez model \eq{ \mathcal L=
F_{\mu\nu}^2 + \lambda D\lambda + \mathcal L_{KK}, } where
$F_{\mu\nu}^2$ is the usual YM curvature and $\lambda$ is the
corresponding majorana spinor. The last term $\mathcal L_{KK}$ is
the lagrangian describing the KK modes and is given by (for recent
discussion see \cite{dorey})
\eq{ \mathcal L_{KK}= (D \phi)^2 +
\Psi D \Psi + V(\lambda, \Psi, \phi), }
where $\phi$ is a scalar,
$\Psi$ is a spinor and V is a potential whose form is not
important in this discussion. So, what Maldacena and Nunez have
done is actually to write a background that represents the strong
coupling regime of the lagrangian
\eq{ \mathcal L= F_{\mu\nu}^2 +
\lambda D \lambda +  (D \phi)^2 + \Psi D \Psi + V(\lambda, \Psi,
\phi) }
which is nothing but an UV completion of $\mathcal N=1$ SYM.

Let us turn now to what is done in \cite{nunez}. In this paper
Gursoy and Nunez provided a continuous set of UV completions
parameterized by $\gamma$. For each value of this parameter we
have difference in the dynamics of the KK modes due to the changes
in the potential $V(\phi, \lambda, \Psi)$ by a dipole deformation.
The conclusions from our computations can be put in two different,
complementary and physically nice ways:

1) In the first type of configurations the strings charged under
$U(1)\times U(1)$ do experience changes in their spectrum and
dynamical relations $E=E(J_1,J_2)$ as the background is deformed.
Hence, those strings "remember" that they are composed out of a
large number of KK modes (which are charged under $U(1)\times
U(1)$ and hence the $\gamma$ deformation changes its dynamics )
and also SYM operators; these are strings of the form
(\ref{opers1}) above. Strings like the one in section 7 (which are
of the form (\ref{opers2})) are made out of {\it only} SYM
operators. Hence they do not see changes in the deformed
background.

2) Our considerations show that the  Gursoy-Nunez backgrounds do
indeed provide different UV completions and that the different
completions are labelled by a continuous parameter $\gamma$. The
different values of $\gamma$ does not change everything, but {\it
just} the UV completion that the supergravity solution is giving
to $\mathcal N=1$ SYM.

Giving support to the Gursoy-Nunez conjecture a reasonable
question arises: how to decouple completely the UV completion?
Certainly this is an open question, because it is not known how to
calculate string theory in RR backgrounds. What will happen if
such a technique is available? One can speculate that the KK modes
will be made infinitely heavy and the spectrum will be just that
of SYM. Then our rotating strings that do depend on $\varphi$ and
$\tilde\varphi$ upon quantization will become very heavy also. The
present situation however do not alow all that and  we have to
live with the KK modes. Since the mass of these KK modes is
similar to the scale of confinement, the relevant and important
problem is to decide whether or not the background we consider is
giving us a result we can trust or not.

As a result of the above discussion, we see that the proposal of
Gursoy and Nunez is to gamma deform the background (which has
completely different behavior in the UV) and compute the same
things in {\it both} backgrounds. If the results depend on
$\gamma$ in the deformed background, the result contains effects
due to KK modes, if not - this is purely gauge theory effects. Our
considerations of rotating strings in different directions, in our
opinion, are giving a very remarkable check of all this picture
described above.

As we pointed out in the previous Section, sometimes it happens that KK modes can cancel in some way the gamma deformations and one can end up with the same result but with KK modes not really decoupled.
Certainly the idea of using gamma deformation of such backgrounds
to separate the contributions from KK modes from the pure gauge
theory effects deserves further study and we will return to this
issue in future investigations. 

\ \\

\textbf{Acknowledgments} We would like to thank Carlos Nunez for
suggesting this problem to us and for his valuable comments. The
work of N.P.B. was partially supported by an EVRIKA foundation
educational award.

\sect {Appendix: A circular string in the background (\ref{L1}).}

Let us consider the following circular string ansatz\footnote{In this appendix we use the notations $X_1=x, X_2=y$.}:

\eq{\begin{array}{l}
x=\,\rho\,\cos\left(\omega_1\,\tau+\alpha(\sigma)\right);\,\,\,\,\psi=\omega_2\,\tau+\beta(\sigma);\,\,t=\kappa\,\tau\\
y=\,\rho\,\sin\left(\omega_1\,\tau+\alpha(\sigma)\right);\,\,\,\,\rho=const;\,\,\,\,r=r(\sigma).
\end{array}\label{L2}}

The Virasoro constraints for this ansatz are:
\begin{equation}
\rho^2\omega_1\,\alpha^{\,\prime}(\sigma)+\frac{\alpha^{\prime}g_{s}N}{4}\,\omega_2\,\beta^{\,\prime}(\sigma)=0,\label{L3}
\end{equation}
\begin{equation}
\rho^2\,{\alpha}^{\,\prime\,2}(\sigma)+\frac{\alpha^{\prime}g_{s}N}{4}\,{\beta}^{\,\prime\,2}(\sigma)+\alpha^{\prime}g_{s}N\,{r}^{\,\prime\,2}(\sigma)+\rho^2\omega_1^2+\frac{\alpha^{\prime}g_{s}N}{4}\,\omega_2^2-\kappa^2=0,\label{L4}
\end{equation}
or
\begin{equation}
\left(\rho^2+\frac{4\rho^4}{A}\,\frac{\omega_1^2}{\omega_2^2}\right)\,{\alpha}^{\,\prime\,2}(\sigma)+A\,{r}^{\,\prime\,2}(\sigma)=B,\label{L5}
\end{equation}
where
$$
A=\alpha^{\prime}\,g_{s}\,N,\,\qquad\,B=\,\kappa^2-\rho^2\omega_1^2-\frac{A}{4}\,\omega_2^2\,>\,0.
$$
The induced world-sheet matric with Virasoro constraints taken
into account has the form:
\begin{equation}
ds^2_{ws}=\,e^{\Phi\left(r(\sigma)\right)}\,B\,\left(-d\tau^2+d\sigma^2\right)\label{L6}
\end{equation}
and then we have
\begin{equation}
\sqrt{-g}\,g^{p\,q}\,\equiv\,\gamma^{p\,q}\,=\,diag(-1,1).\label{L7}
\end{equation}

The relevant Polyakov action for this geometry is:
\begin{multline}
S=-\frac{1}{4\pi}\int\,d\tau\,d\sigma\,e^{\Phi\left(r(\sigma)\right)}\,[\kappa^2-\rho^2\omega_1^2-\frac{A}{4}\,\omega_2^2+\\
+\rho^2\,{\alpha}^{\,\prime\,2}(\sigma)+\frac{A}{4}\,{\beta}^{\,\prime\,2}(\sigma)+A\,{r}^{\,\prime\,2}(\sigma)].\label{L8}
\end{multline}

The equations of motion for $\,\alpha(\sigma)\,$ and
$\,\beta(\sigma)\,$ are the identical:
\begin{equation}
\alpha^{\,\prime\prime}(\sigma)+\frac{d\Phi(r)}{dr}\,r^{\,\prime}(\sigma)\,\alpha^{\,\prime}(\sigma)=0,\label{L9}
\end{equation}
\begin{equation}
\beta^{\,\prime\prime}(\sigma)+\frac{d\Phi(r)}{dr}\,r^{\,\prime}(\sigma)\,\beta^{\,\prime}(\sigma)=0.\label{L10}
\end{equation}
Integrating them once we obtain
\begin{align}
\frac{d}{d\sigma}\,{\alpha}^{\,\prime\,2}(\sigma)+2\,\frac{d}{d\sigma}\,\Phi(r)\,{\alpha}^{\,\prime\,2}(\sigma)=0,\label{L11}\\
{\alpha}^{\,\prime\,2}(\sigma)\,=\,C_1\,e^{-2\Phi(r)},\label{L12}\\
{\beta}^{\,\prime\,2}(\sigma)\,=\,C_2\,e^{-2\Phi(r)}.\label{L13}
\end{align}
One can check that the Virasoro constraint \eqref{L3} are
satisfied.

For $\,r(\sigma)\,$ we have the following equation of motion:
\begin{equation}
2A\,r^{\,\prime\prime}(\sigma)+A\,\frac{d\Phi(r)}{dr}\,r^{\,\prime\,2}(\sigma)-\frac{d\Phi(r)}{dr}\,\left[B+\rho^2\,{\alpha}^{\,\prime\,2}(\sigma)+\frac{A}{4}\,{\beta}^{\,\prime\,2}(\sigma)\right]=0.\label{L14}
\end{equation}
Using the Virasoro constraint \eqref{L4} we obtain for
$\,r(\sigma)\,$ the equation
\begin{equation}
A\,r^{\,\prime\prime}(\sigma)+A\,\frac{d\Phi(r)}{dr}\,r^{\,\prime\,2}(\sigma)-\frac{d\Phi(r)}{dr}\,B=0.\label{L15}
\end{equation}
One can easily integrate this equation as follow:
$$A\,\frac{d}{d\sigma}r^{\,\prime\,2}(\sigma)+2\frac{d\Phi(r)}{dr}\,r^{\,\prime}(\sigma)\left[A\,r^{\,\prime\,2}(\sigma)-B\right]=0$$
$$\frac{d}{d\sigma}\left[A\,r^{\,\prime\,2}(\sigma)-B\right]+2\left[A\,r^{\,\prime\,2}(\sigma)-B\right]\frac{d}{d\sigma}\Phi(r)=0$$
$$\frac{1}{\left[A\,r^{\,\prime\,2}(\sigma)-B\right]}\frac{d}{d\sigma}\left[A\,r^{\,\prime\,2}(\sigma)-B\right]=-2\frac{d}{d\sigma}\Phi(r)$$
$$|\,A\,r^{\,\prime\,2}(\sigma)-B\,|\,=\,C\,e^{-2\Phi(r)}$$
In order the resulting solution to be periodic in $\sigma$ one
should take the constant $C$ negative.
At the end of the day we find
\begin{equation}
r^{\,\prime\,2}(\sigma)=\,\frac{B}{A}-\frac{C}{A}\,e^{-2\Phi(r)},\label{L16}
\end{equation}
where $\,\frac{B}{C}\,<\,e^{-2\Phi_0}\,$.
 This equation can be rewritten as:
\begin{equation}
d\sigma\,=\,\frac{1}{\sqrt{\frac{B}{A}-\frac{C}{A}\,e^{-2\Phi(r)}}}\,dr.\label{L17}
\end{equation}
This relation is useful for the computation of the conserved
charges in our problem (the energy and angular momenta ) which can
be written in the form:
\begin{align}
E\,=\,\frac{\kappa}{2\pi}\int\limits_0^{2\pi}\,e^{\Phi(r)}\,d\sigma,\label{L18}\\
J_1\,=\,\frac{\rho^2\omega_1}{2\pi}\int\limits_0^{2\pi}\,e^{\Phi(r)}\,d\sigma,\label{L19}\\
J_2\,=\,\frac{(A/4)\,\omega_2}{2\pi}\int\limits_0^{2\pi}\,e^{\Phi(r)}\,d\sigma.\label{L20}
\end{align}
Therefore, the final expression for the energy takes the form
\begin{equation}
E\,=\,\frac{\kappa}{\rho^2\omega_1+\frac{A}{4}\omega_2}\,(J_1+J_2).\label{L21}
\end{equation}

We want to emphasize that this result is exact and does not depend
on the deformation parameter.

To complete the analysis, we will present here the cases of the
limits of short and long strings. In the first case we have
$\,r\longrightarrow0\,$. In this limit the dilaton field has the
following behavior:
\begin{equation}
e^{-2\Phi(r)}\,\approx\,e^{-2\Phi_0}\,\left[1-\frac{8}{9}\,r^2+\frac{224}{405}\,r^4\right].\label{L22}
\end{equation}
Substituting this approximation into the equation \eqref{L16} we
get:
\begin{equation}
r^{\,\prime\,2}(\sigma)=\,\frac{B}{A}-\frac{C}{A}\,e^{-2\Phi_0}\,\left[1-\frac{8}{9}\,r^2+\frac{224}{405}\,r^4\right],\label{L23}
\end{equation}
or
\begin{equation}
r^{\,\prime\,2}(\sigma)=\frac{224}{405}\,\frac{C}{A}\,e^{-2\Phi_0}\,(r^2_2-r^2)\,(r^2-r^2_1),\label{L24}
\end{equation}
where
\begin{align}
r_1^2=\frac{45}{56}\,\left[1-\frac{3}{\sqrt{5}}\,\sqrt{\frac{14}{9}\,\frac{B}{C}\,e^{2\Phi_0}-1}\,\right]\,<\,1,\label{L25}\\
r_2^2=\frac{45}{56}\,\left[1+\frac{3}{\sqrt{5}}\,\sqrt{\frac{14}{9}\,\frac{B}{C}\,e^{2\Phi_0}-1}\,\right]\,<\,1\label{L26}
\end{align}
are the turning points of the short string,
$\,0<r_1^2<r^2<r_2^2<1\,$ and
$\,\frac{9}{14}\,<\,\frac{B}{C}\,e^{2\Phi_0}<\,1\,$.

The $2\pi$ periodic solution of \eqref{L24} is:
\begin{equation}
r(\sigma)\,=\,r_2\,\mathbf{dn}\,\left[\frac{\mathbf{K}}{\pi}\,\left(\frac{\pi}{2}-\sigma\right)\,|\,\frac{r_2^2-r_1^2}{r_2^2}\right],\label{L27}
\end{equation}
where
$\,C\,=\,(\frac{9\mathbf{K}}{4\pi\,r_2})^2\,\frac{5}{14}\,A\,e^{2\Phi_0}\,$.

For energy and spins in short string case we have:
\begin{multline}
E\,=\,\frac{\kappa}{2\pi}\int\limits_0^{2\pi}\,e^{\Phi(r)}\,d\sigma\,=2\,\frac{\kappa}{2\pi}\int\limits_{r_1}^{r_2}\,\frac{e^{\Phi(r)}}{\sqrt{\,\frac{B}{A}-\frac{C}{A}\,e^{-2\Phi(r)}}}\,dr\approx\\
\approx\frac{\kappa\,r_2\,e^{\Phi_0}}{\mathbf{K}}\,\int\limits_{r_1}^{r_2}\,\frac{1+\frac{4}{9}\,r^2+\frac{8}{405}\,r^4}{\sqrt{\,(r^2_2-r^2)\,(r^2-r^2_1)}}\,dr=\\
=e^{\Phi_0}\,\kappa+\frac{4}{9}\,e^{\Phi_0}\,\kappa\,r_2^2\,\frac{\mathbf{E}}{\mathbf{K}}+\frac{8}{405}\,\frac{e^{\Phi_0}\,\kappa\,r_2^2}{3\mathbf{K}}\,[2(r_1^2+r_2^2)\,\mathbf{E}-r_1^2\,\mathbf{K}],\label{L28}
\end{multline}
\begin{align}
J_1\,=\,e^{\Phi_0}\,\rho^2\omega_1\,\left[1+\frac{4}{9}\,r_2^2\,\frac{\mathbf{E}}{\mathbf{K}}+\frac{8}{405}\,\frac{r_2^2}{3\mathbf{K}}\,[2(r_1^2+r_2^2)\,\mathbf{E}-r_1^2\,\mathbf{K}]\right],\label{L29}\\
J_2\,=\,e^{\Phi_0}\,\frac{A}{4}\omega_2\,\left[1+\frac{4}{9}\,r_2^2\,\frac{\mathbf{E}}{\mathbf{K}}+\frac{8}{405}\,\frac{r_2^2}{3\mathbf{K}}\,[2(r_1^2+r_2^2)\,\mathbf{E}-r_1^2\,\mathbf{K}]\right].\label{L30}
\end{align}

In long string case (i.e. $\,r\longrightarrow\infty\,$) the
dilaton field has the following behavior:
\begin{equation}
e^{-2\Phi(r)}\,\approx\,\frac{4r^{1/2}}{e^{2r}}.\label{L31}
\end{equation}
Using this approximation in the equation \eqref{L16}we get
\begin{equation}
r^{\,\prime\,2}(\sigma)=\,\frac{B}{A}-\frac{C}{A}\,\frac{4r^{1/2}}{e^{2r}}.\label{L32}
\end{equation}
The turning point $r_0$ satisfies the equality
\begin{equation}
\frac{4r_0^{1/2}}{e^{2r_0}}=\frac{B}{C}.\label{L33}
\end{equation}

For energy and spins we find:
\begin{multline}
E\,=\,\frac{\kappa}{\pi}\,\sqrt{\frac{A}{C}}\int\limits_0^{r_0}\,\frac{e^r\,r^{-1/4}}{\sqrt{\frac{B}{C}-\frac{4r^{1/2}}{e^{2r}}}}\,dr\approx\\
\approx\,\frac{\kappa}{\pi}\,\sqrt{\frac{A}{B}}\,r_0^{3/4}\,\sqrt{2}\,\frac{\Gamma(1/2)\,\Gamma(1)}{\Gamma(3/2)}\,{}_1\mathrm{F}_1
\left[1,\,3/2,\,r_0\right]\approx\,\frac{\kappa}{\sqrt{\pi}}\,\sqrt{\frac{A}{B}}\,\sqrt{2}\,r_0^{1/4}\,e^{r_0},\label{L34}
\end{multline}
\begin{align}
J_1\,\approx\,\frac{\rho^2\omega_1}{\sqrt{\pi}}\,\sqrt{\frac{A}{B}}\,\sqrt{2}\,r_0^{1/4}\,e^{r_0},\label{L35}\\
J_2\,\approx\,\frac{A\,\omega_2}{4\sqrt{\pi}}\,\sqrt{\frac{A}{B}}\,\sqrt{2}\,r_0^{1/4}\,e^{r_0}.\label{L36}
\end{align}

The expressions we found for the limiting cases of short and long
strings are consistent with the general result (\ref{L21}) we
already found.

The string configurations we studied in this Appendix are somhow trivial. This i because the circular string is "spinning into itselfs" and looks like point particle circling on the circle with radius $\rho$. Nevertheless, it is good to have explicit solutions supporting the conjecture of separating the KK modes. 

At the end, one can conclude from the considerations in this Appendix that there
exist cases in which the KK modes are lifted by the deformation of
the geometry and the obtained results are purely gauge theory
effects.

\end{document}